\documentclass[
prb, 11pt, letterpaper,
superscriptaddress, floatfix,
nofootinbib, notitlepage
]{revtex4-1}

\usepackage{amsmath, amsfonts, amssymb}
\usepackage{bm} 

\usepackage{graphicx}
\usepackage{siunitx}
\usepackage{caption}
\usepackage{xcolor}
\usepackage{float}

\usepackage{textcomp}
\usepackage{setspace}
\usepackage[colorlinks, citecolor={blue}, linkcolor={black}]{hyperref}

\setcitestyle{super}

\newcommand*{\vdot}{\bm{\cdot}} 
\newcommand*{\vb}[1]{\bm{#1}} 

\newcommand{\rf}[1]{\textcolor{magenta}{[REF]}}

\makeatletter
\newcommand{\keepEDref}[2]{%
  \@namedef{r@#1}{{#2}{}{}{}{}}%
}

\keepEDref{fig:lin-T-filling}{1}
\keepEDref{fig:field_dependence_tunneling_transport}{2}
\keepEDref{fig:B-calibration}{3}
\keepEDref{fig:Full-gap-no-nema}{4}
\keepEDref{fig:Parameter-tuning}{5}
\keepEDref{fig:Landau_fans}{6}
\keepEDref{fig:d2IdVdB-anisotropy}{7}
\keepEDref{fig:strain-anisotropy}{8}
\makeatother

\begin{document}
\title{Nodal nematic superconductivity in moiré graphene}

\author{Shuwen Sun}
\affiliation{Department of Physics, Massachusetts Institute of Technology, Cambridge, Massachusetts 02139, USA}

\author{Grgur Palle}
\affiliation{Department of Physics, The Grainger College of Engineering, University of Illinois Urbana-Champaign, Urbana, Illinois 61801, USA}
\affiliation{Anthony J. Leggett Institute for Condensed Matter Theory, The Grainger College of Engineering, University of Illinois Urbana-Champaign, Urbana, Illinois 61801, USA}

\author{Kenji Watanabe}
\author{Takashi Taniguchi}
\affiliation{National Institute for Materials Science, Namiki 1-1, Tsukuba, Ibaraki 305-0044, Japan}

\author{Rafael M. Fernandes}
\affiliation{Department of Physics, The Grainger College of Engineering, University of Illinois Urbana-Champaign, Urbana, Illinois 61801, USA}
\affiliation{Anthony J. Leggett Institute for Condensed Matter Theory, The Grainger College of Engineering, University of Illinois Urbana-Champaign, Urbana, Illinois 61801, USA}

\author{Pablo Jarillo-Herrero}
\email{parkjane@mit.edu}\email{pjarillo@mit.edu}
\affiliation{Department of Physics, Massachusetts Institute of Technology, Cambridge, Massachusetts 02139, USA}

\author{Jeong Min Park}
\email{parkjane@mit.edu}\email{pjarillo@mit.edu}
\affiliation{Department of Physics, Massachusetts Institute of Technology, Cambridge, Massachusetts 02139, USA}
\affiliation{Department of Physics, Princeton University, Princeton, New Jersey 08544, USA}
\maketitle

\textbf{Unconventional superconductivity can emerge in various forms in strongly correlated systems\cite{Sigrist1991}. Historically, the most extensively studied examples include nodal and triplet pairing, which were experimentally detected in cuprates\cite{keimer2015quantum} and superfluid helium-3\cite{vollhardt2013superfluid}, respectively. Beyond these canonical cases, there are also nematic superconductors, in which the rotational symmetry of the lattice is broken in the superconducting state. While there are several examples of nematic superconductors whose pairing emerges inside a preexisting nematic normal state\cite{Howald2003,Kim2008,Wu2017,Sprau2017,Nag2025}, experimental evidence for superconducting states that intrinsically break rotational symmetry on their own has been exceedingly scarce. Here, we report the experimental realization and identification of an intrinsic multi-component nodal nematic superconducting state in magic-angle twisted trilayer graphene (MATTG). By integrating tunneling spectroscopy with quantum transport measurements, we directly probe the symmetry of low-energy superconducting quasiparticles. Deep inside the superconducting state, we observe a pronounced two-fold anisotropy in the zero-bias tunneling conductance under an in-plane magnetic field, which is strongly suppressed together with superconductivity as the temperature or perpendicular magnetic field is increased. Remarkably, the major axis of anisotropy rotates by up to $\SI{90}{^{\circ}}$ across the superconducting dome, indicating an intrinsic, field- and density-tunable symmetry of the low-energy excitations. These behaviors imply a nodal superconducting state in MATTG that spontaneously breaks the six-fold rotational symmetry of the moir\'e superlattice. Our results establish moiré graphene as a versatile platform for realizing and interrogating nodal nematic superconductivity, opening a new avenue for exploring unconventional symmetry-breaking superconducting states in two-dimensional materials.}

Superconductivity in which electronic interactions stabilize a pairing state with unconventional gaps has been experimentally established only in a limited number of material systems. Notable examples include the cuprate superconductors\cite{keimer2015quantum}, where gap nodes give rise to low-energy quasiparticles, and iron-based superconductors \cite{Fernandes2022}, which generally display nodeless gaps of opposite signs on different Fermi surfaces. Interactions have also been proposed to generate superconducting states that spontaneously break the rotational symmetry of the underlying lattice, leading to nematic superconductivity\cite{Fernandes2013, Fu2014, Hecker2018, Kozii2019, Chichinadze2020, Fernandes_Fu, Jian2021}. Such intrinsic nematic superconductors necessarily require a multi-component pairing, and are thus fundamentally distinct from another type of nematic superconductivity in which single-component pairing develops in an already nematic normal state \cite{Dodaro2018},  as it has been observed in certain cuprates\cite{Howald2003,Kim2008,Wu2017} and iron chalcogenides \cite{Sprau2017,Nag2025}. Meanwhile, experimental evidence for multi-component nematic superconductivity remains rare, with copper-intercalated bismuth selenide being the most widely discussed candidate\cite{matano_spin-rotation_2016,yonezawa_thermodynamic_2017}. 
While the pairing in this compound is believed to be fully gapped, the interplay between nodal superconducting quasiparticles and nematicity has been widely studied theoretically, and shown to be connected to unique quantum critical and topological behavior\cite{Vojta2000, Wang_Fernandes2021, Yu_Sarma2023, Song2024}. A multi-component nodal nematic superconductor would thus open new routes to realize such exotic phenomena. Challenges in identifying intrinsic nodal nematic superconductors arise not only from the fact that these two phenomena, nodality and nematicity, are relatively rare, but also because typical experimental probes employed to assess nematic pairing and gap nodes operate in opposite regimes: close to $T_c$ and at temperatures much smaller than $T_c$, respectively.

Beyond bulk materials, moiré graphene has emerged as a promising platform for exploring interaction-driven instabilities and superconductivity\cite{cao_correlated_2018,cao_unconventional_2018,andrei_marvels_2021,park_tunable_2021,hao_electric_2021,park_robust_2022,zhang_promotion_2022}. The tunability with gate voltages and twist angle enables an unprecedented level of control, which can prevent some of the ambiguities in establishing unconventional pairing states in bulk materials. Indeed, a growing body of experimental evidence points toward unconventional superconductivity in these systems\cite{cao_pauli-limit_2021,oh_evidence_2021,kim_evidence_2022,banerjee_superfluid_2025,tanaka_superfluid_2025}. In magic-angle twisted trilayer graphene (MATTG), tunneling spectroscopy has revealed a distinct double V-shaped gap structure\cite{park_simultaneous_2025, kim2026resolving}, comprising a low-energy superconducting gap and a higher-energy pseudogap-like feature, together with additional behaviors of the zero-bias tunneling conductance that are consistent with the presence of nodes, such as a linear dependence on temperature and a square-root dependence on the out-of-plane magnetic field (Volovik effect)\cite{volovik_superconductivity_1993, moler1994magnetic}. Furthermore, in both twisted bilayer graphene and trilayer graphene, transport measurements have suggested rotational symmetry-breaking in the normal-to-superconducting transition and in the normal state\cite{cao_nematicity_2021, zhang2025angular}. Taken together, these observations identify magic-angle graphene as a potential candidate for hosting intrinsic nodal nematic superconductivity. However, direct experimental access to the rotational symmetry deep inside the superconducting state is yet to be realized.

In this article, we report evidence for a nodal, two-fold nematic superconducting pairing symmetry in mirror-symmetric magic-angle twisted trilayer graphene. To overcome the limitations of bulk probes in vdW heterostructures, we develop a novel platform that integrates transport measurements and tunneling spectroscopy within the same device, enabling a direct density of states (DOS) based characterization of rotational symmetry while independently tracking the onset of superconductivity\cite{park_simultaneous_2025}. Using this platform, we investigate the in-plane anisotropy of the zero-bias tunneling conductance in the superconducting state. We perform systematic measurements by varying the magnetic field, both along in-plane and out-of-plane directions, and temperature, and establish a direct correlation between the emergence of the tunneling conductance anisotropy and the onset of superconductivity in MATTG. We further explore the doping dependence of the zero-bias tunneling anisotropy, thereby revealing a rotating major axis, as well as a varying anisotropy ratio, across the superconducting dome. These results imply the existence of nodes that spontaneously break the six-fold rotational symmetry of MATTG. To confirm this conclusion and further constrain the type of pairing state that can give rise to an anisotropic zero-bias tunneling conductance, we also perform a general phenomenological analysis. We find that, in the presence of mirror symmetry, the most favorable state is a nodal nematic superconductor with a dominant triplet component characterized by an in-plane d-vector.

\begin{figure}
\includegraphics[width=1\textwidth]{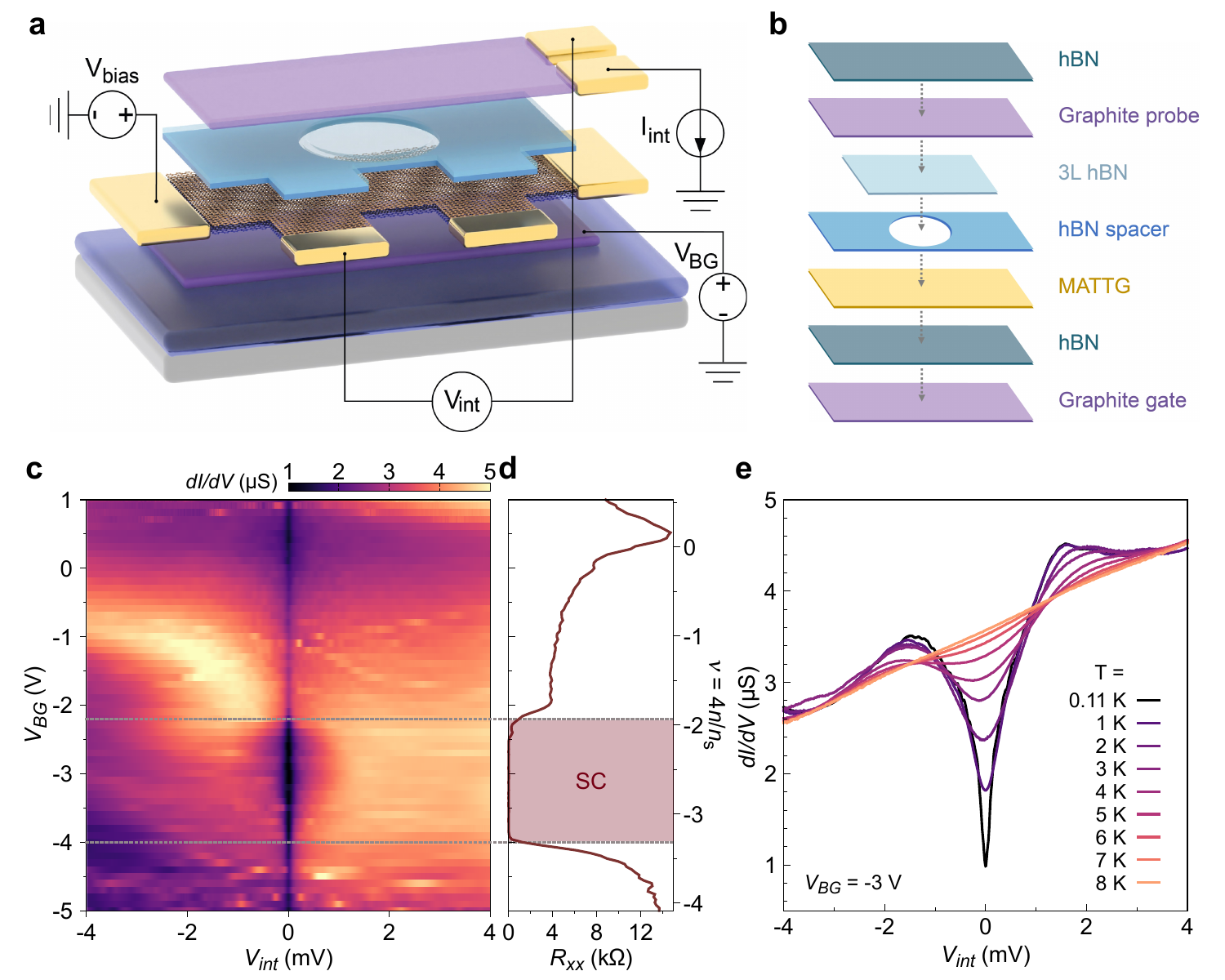}
\caption{\small Characterization of the simultaneous transport and tunneling device. (a) Schematic illustration of the device structure. The magic-angle twisted trilayer graphene (MATTG) is coupled to the graphite finger probe through a trilayer hBN tunnel barrier. The tunneling is restricted to a circular region inside the Hall bar device, defined by an additional hBN spacing layer with a hole. (b) The sequence of each layer in the device. (c) The tunneling conductance $dI/dV$ as a function of interlayer bias voltage $V_{int}$ and bottom gate voltage $V_{BG}$ at $T=\SI{105}{\milli\kelvin}$. (d) Transport resistance $R_{xx}$ of the MATTG measured as a function of $V_{BG}$. The zero-resistance region, colored in red, in the transport measurement corresponds directly to the region where the V-shaped superconducting gaps exist in the tunneling spectra. The gray dashed line is a guide to the eye for this correspondence. (e) Temperature-dependent tunneling spectra at around optimal doping $V_{BG}=\SI{-3}{\volt}$ from the base temperature up to $\SI{8}{\kelvin}$. The superconducting critical temperature is $T_{c}=\SI{1.1}{\kelvin}$ according to transport measurements. The coexistence of two V-shaped gaps at distinct energy scales is evident, with the smaller-energy superconducting gap vanishing as temperature increases toward the normal state (see also in Extended Data Fig.~\ref*{fig:lin-T-filling}) and the higher-energy pseudogap-like feature persisting until $\SI{8}{\kelvin}$.}
\label{fig:device-character}
\end{figure}

Figure~\ref{fig:device-character}a shows the simultaneous transport and tunneling measurement scheme, where MATTG is tunnel-coupled to the graphite finger probe via a trilayer hexagonal boron nitride (hBN). We insert a few-nanometer-thick hBN spacing layer with a lithographically defined hole between the MATTG and the trilayer hBN to confine the tunneling region to the center of the Hall bar device, facilitating a direct comparison between the transport and tunneling measurements. A bottom graphite gate is implemented to control the electrostatic doping of MATTG. The layered structure of the device is schematically illustrated in Fig.~\ref{fig:device-character}b. Figure~\ref{fig:device-character}c shows the tunneling conductance, $dI/dV$, as a function of the interlayer bias voltage $V_{int}$ and bottom gate voltage $V_{BG}$, measured at base temperature $T=\SI{105}{\milli\kelvin}$. The broad peak starting from $V_{BG}=\SI{-1}{\volt}$ on the negative bias side, whose energy and width depend on the carrier density, results from the interaction-driven broadening of the flat bands. There is a strong depletion of the DOS (gapped region) in the doping range $V_{BG}=$\SIrange{-4}{-2.2}{\volt}. Simultaneous four-probe transport measurements reveal dissipationless transport with zero resistance in the same doping range (red-colored region in Fig.~\ref{fig:device-character}d). Such correspondence indicates that the gap observed with tunneling spectroscopy and the superconductivity in this device have a direct relationship, as schematically shown by the gray dashed lines in Fig.~\ref{fig:device-character}c and~\ref{fig:device-character}d, and consistent with our previous measurements in a similar device geometry\cite{park_simultaneous_2025}. Outside the superconducting dome, a much weaker suppression of the DOS persists over a broad gate-voltage range. It may be potentially associated with electron-electron interactions and localization of electrons by moiré disorder\cite{xie2019spectroscopic}, which is distinct from the low-energy superconducting gap discussed in this work.

The temperature dependence of $dI/dV$ versus $V_{int}$ around optimal doping is shown in Fig.~\ref{fig:device-character}e. At base temperature, we observe in a single tunneling spectrum the coexistence of a sharp, V-shaped gap at low energies ($\sim \SI{0.1}{meV}$), and a distinct broader gap at higher energies ($\sim \SI{1}{meV}$). As the temperature approaches and exceeds the transport critical temperature $T_c$, the low-energy superconducting gap is rapidly suppressed and disappears near the onset temperature $T_{onset}$ where superconductivity is fully suppressed (see Methods for definition and Extended Data Fig.~\ref*{fig:lin-T-filling}). The high-energy gap (HG), however, remains nearly unperturbed until about $\SI{2}{\kelvin}$, and persists up to $\SI{8}{\kelvin}$. Such behaviors are consistent with previous observations\cite{park_simultaneous_2025} and show that the low-energy V-shaped spectrum corresponds to a nodal superconducting gap in MATTG. We note that other previously reported evidence for nodal superconductivity\cite{park_simultaneous_2025} is also reproduced in this device, including the linear temperature-dependent gap filling behavior and the Volovik effect at low temperature and perpendicular magnetic field (see Extended Data Fig.~\ref*{fig:lin-T-filling} and~\ref*{fig:field_dependence_tunneling_transport}).

In two-dimensional hexagonal systems, the superconducting states can have different nodal gap structures\cite{Sigrist1991}. One way of differentiating between them is to investigate whether the six-fold rotational symmetry of the normal state is broken or preserved in the superconducting state. For instance, time-reversal symmetry-preserving $p$-wave and $d$-wave pairing states break this symmetry, whereas $f$-wave and $g$-wave states preserve it. To probe the rotational symmetry of a superconducting state, one often relies on the anisotropy of the transport in-plane critical magnetic field\cite{palstra1988angular, li2017nematic, hamill2021two}. Despite the important insight offered by this quantity, it probes a regime in which superconducting long-range order is weak and fluctuations dominate. Moreover, as extracted from the resistivity, it can be affected by several factors, from inhomogeneities\cite{Vaglio1993} to details of the Fermi surface\cite{Wang2023}. Thus, to directly assess the anisotropy deep inside the superconducting state, here we measure the dependence of the tunneling spectroscopy on the direction of the in-plane magnetic field. Although the absence of a two-fold anisotropy does not necessarily imply that the superconductivity is non-nematic, for instance, if the effect is small and below experimental resolution, its presence necessarily requires a nematic superconducting state. As we discuss below, in MATTG there are, in principle, two different mechanisms by which the in-plane field can couple to the Cooper pairs and thus affect the zero-bias conductance: mirror symmetry-breaking and spin-orbit coupling. Our conclusion of a nodal nematic superconductor, however, does not depend on which mechanism is at play.

\begin{figure}
\includegraphics[width=1\textwidth]{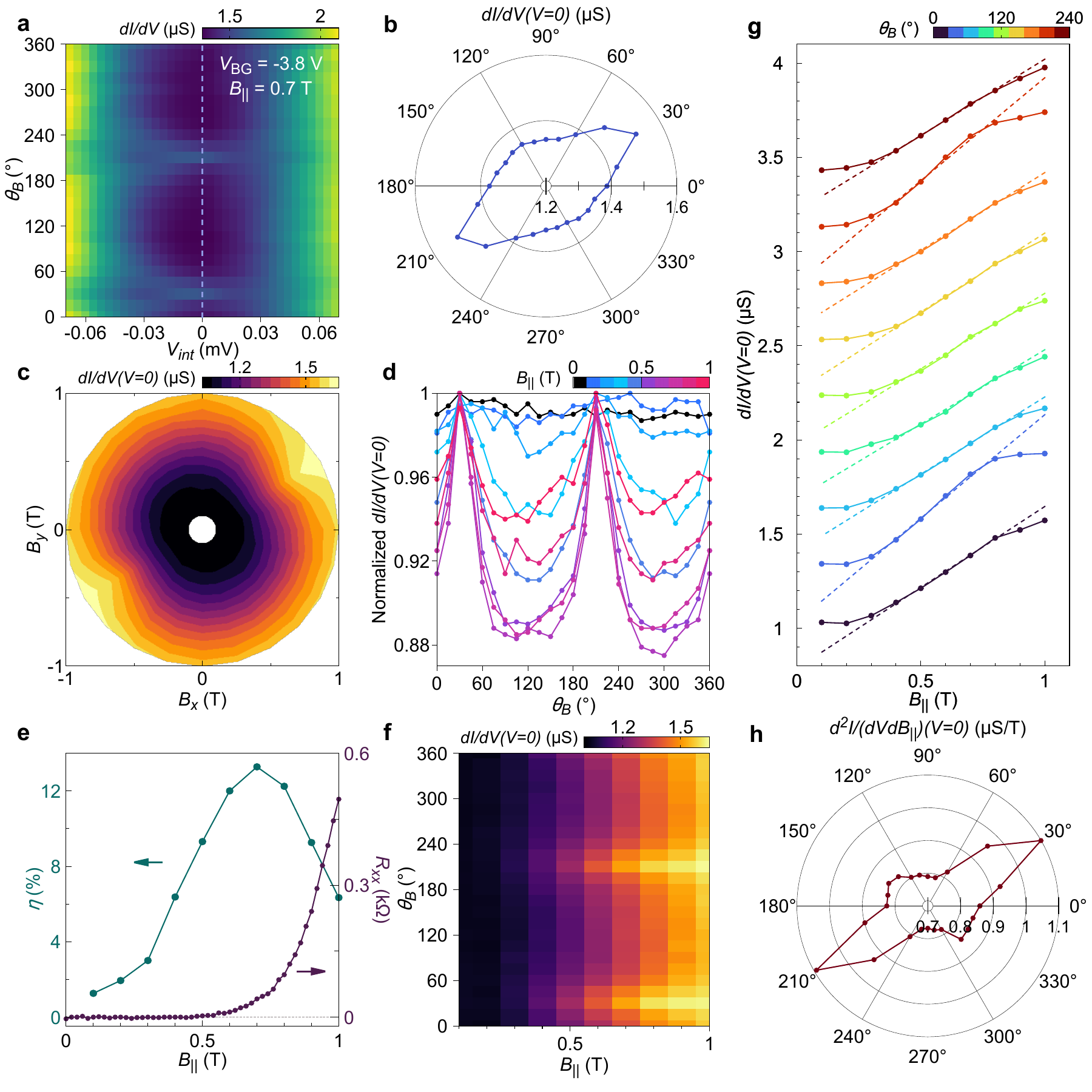}
\caption{\small Tunneling spectroscopy anisotropy. (a) Tunneling conductance $dI/dV$ as a function of $V_{int}$ and the in-plane magnetic field rotation angle $\theta_{B}$ at the overdoped side of the superconducting dome $V_{BG}=\SI{-3.8}{\volt}$ and for the in-plane magnetic field magnitude $B_{\parallel}=\SI{0.7}{\tesla}$. The dashed light purple line tracks $V_{int}=\SI{0}{\volt}$, around which the tunneling conductance exhibits a clear two-fold anisotropy as a function of $\theta_{B}$. (b) Polar plot of the angular-dependent zero-bias tunneling conductance $dI/dV(V=0)$ as denoted by the dashed line in (a). The radial axis range is chosen to visually emphasize the anisotropy of $dI/dV(V=0)$, with concentric circles serving as reference markers.  (c) Polar map of the anisotropic response of $dI/dV(V=0)$ with respect to $B_{\parallel}$ and $\theta_{B}$. (d) $dI/dV(V=0)$, normalized by its maximum value at each $B_{\parallel}$ value, as a function of $\theta_{B}$ for different in-plane magnetic field magnitudes. (e) Anisotropy ratio $\eta$ versus $B_{\parallel}$ (green, left axis) up to $\SI{1}{\tesla}$. The maximum anisotropy takes place at $B_{\parallel}\sim\SI{0.7}{\tesla}$. The figure also shows the dependence of $R_{xx}$ on $B_{\parallel}$ (purple, right axis) for comparison. (f) $dI/dV(V=0)$ as a function of $B_{\parallel}$ and $\theta_{B}$, showing a prominent two-fold anisotropy, especially in the intermediate to high $B_{\parallel}$ range. (g) $B_{\parallel}$ dependence of $dI/dV(V=0)$ across different $\theta_{B}$ values, which are shifted vertically for clarity. The dashed lines are linear fits of the data points in the in-plane magnetic field range \SIrange{0.4}{0.8}{\tesla}. The angular-dependent slopes of the linear fits are extracted and plotted in (h), indicating a distinct two-fold anisotropy in the same direction as the $dI/dV(V=0)$ anisotropy. All data were taken at $T=\SI{105}{\milli\kelvin}$ and $V_{BG}=\SI{-3.8}{\volt}$.}
\label{fig:exp-anisotropy}
\end{figure}

\begin{figure}
\includegraphics[width=1\textwidth]{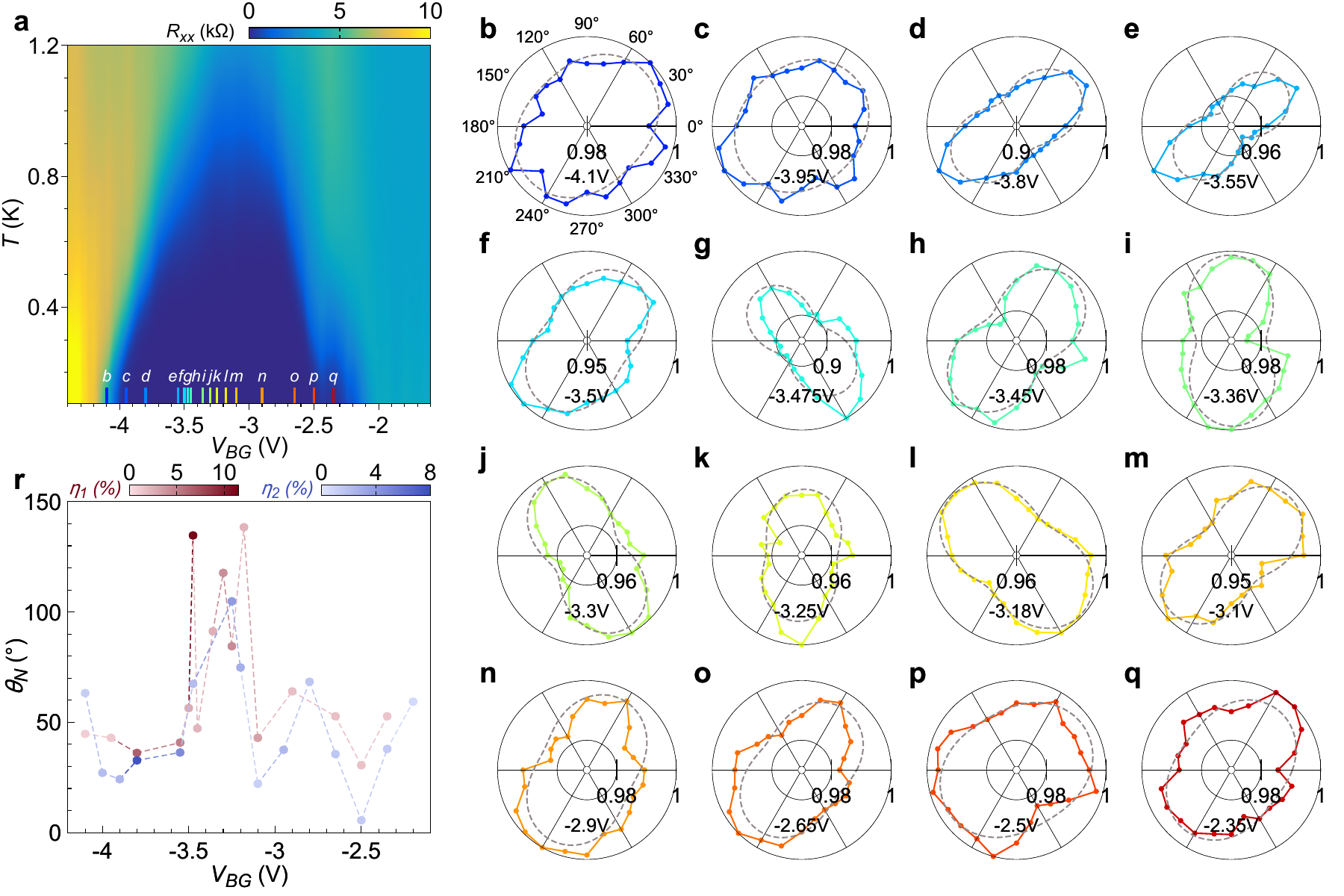}
\caption{\small Gate-tunable tunneling conductance anisotropy across the superconducting dome. (a) $R_{xx}$ as a function of $V_{BG}$ and $T$ in the hole-doped superconducting regime. (b-q) Polar plots of the normalized $dI/dV(V=0)$ as a function of $\theta_{B}$ across the hole-doped superconducting dome, measured at $T=\SI{300}{\milli\kelvin}$ and $B_{\parallel}=\SI{0.7}{\tesla}$. The gray dashed lines represent fits using an ellipse function (see Methods). The radial axis range in each polar plot is individually chosen to reflect the varying magnitudes of anisotropy at different doping levels, with the concentric circles serving as reference markers for the radial axis. Note that the peanut appearance of the curve is related to the selected radial axis range. The $V_{BG}$ values at which (b-q) are measured correspond to the labels at the bottom of (a). (r) $dI/dV(V=0)$ exhibits pronounced anisotropy across the entire doping range. The major axis of the anisotropy $\theta_{N}$ evolves as the carrier density is tuned from the overdoped regime (b) to the underdoped regime (q). Simultaneously, the anisotropy ratio $\eta$ changes in magnitude, with the opacity indicating the magnitude of the anisotropy. $\eta_{1}$ and $\eta_{2}$, along with their corresponding $\theta_{N}$, are extracted from results measured using two different pairs of contacts ($\eta_{1}$ corresponds to the data shown in (b-q), while $\eta_{2}$ is measured using a separate pair of contacts). The magnitude of $\eta_{1}$ and $\eta_{2}$, as well as their $\theta_{N}$, follow a similar trend as a function of $V_{BG}$.}
\label{fig:anisotropy-across-dome}
\end{figure}

We apply $\vb{B}_{\parallel} = B_{\parallel} (\cos \theta_B, \sin \theta_B)$, a magnetic field parallel to the horizontal plane of the junction, and investigate the angular dependence of the tunneling conductance on the field orientation $\theta_{B}$. We compensate for the possible tilting of the sample by applying a small calibrated perpendicular magnetic field $B_{\perp}$, ensuring that the $B_{\perp}$ component is smaller than $\SI{1}{\milli\tesla}$ at each $B_{\parallel}$ and $\theta_{B}$ data point; see Methods and Extended Data Fig.~\ref*{fig:B-calibration} for the detailed field calibration process and representative calibration data. Figure~\ref{fig:exp-anisotropy}a shows a 2D map of $dI/dV$ versus $\theta_{B}$ for a narrow range of $V_{int}$ around zero bias voltage, measured at $V_{BG}=\SI{-3.8}{\volt}$ in the over-doped regime at $B_{\parallel}=\SI{0.7}{\tesla}$, where the superconducting anisotropy is prominent (see Fig.~\ref{fig:anisotropy-across-dome} for detailed characterization). For $dI/dV$ at finite $V_{int}$ away from zero bias, the signal is nearly independent of $\theta_{B}$. However, near zero bias, $dI/dV$ exhibits a pronounced two-fold anisotropy as a function of $\theta_{B}$, breaking the six-fold rotational symmetry of MATTG. We present the $\theta_{B}$ dependence of $dI/dV(V=0)$ as a polar plot in Fig.~\ref{fig:exp-anisotropy}b, extracted along the light purple dashed line in Fig.~\ref{fig:exp-anisotropy}a. The major axis of the anisotropy lies along the direction of $\SI{30}{^{\circ}}$ with respect to the positive $x$-axis of the vector magnet. 

To quantify the magnitude of the tunneling conductance anisotropy at a given $B_{\parallel}$, we define and extract the anisotropy ratio $\eta \equiv 2\frac{dI/dV(V=0)_{max}-dI/dV(V=0)_{min}}{dI/dV(V=0)_{max}+dI/dV(V=0)_{min}}$, where $dI/dV(V=0)_{max/min}$ denotes the maximum/minimum zero-bias tunneling conductance, yielding a value of approximately $\SI{13.3}{\%}$ at $B_{\parallel}=\SI{0.7}{\tesla}$. The two-fold anisotropy is also clearly manifested in the polar map of the in-plane magnetic field response of $dI/dV(V=0)$, as shown in Fig.~\ref{fig:exp-anisotropy}c. The major axis of the anisotropy remains along the same direction as the field magnitude changes. To further analyze the $B_{\parallel}$-dependent behavior of the tunneling conductance anisotropy, in Fig.~\ref{fig:exp-anisotropy}d we plot line cuts of the normalized zero-bias conductance, $dI/dV(V=0)$ rescaled by its maximum value $dI/dV(V=0)_{max}$ at each $B_{\parallel}$, versus $\theta_{B}$ for various magnitudes of $B_{\parallel}$. This plot reveals that the anisotropy magnitude first increases in the low-field range, reaching the largest anisotropy at around $\SI{0.7}{\tesla}$, then gradually decreases as the field further increases up to $\SI{1}{\tesla}$. To better illustrate this behavior, we obtain the anisotropy ratio $\eta$ as a function of $B_{\parallel}$ from Fig.~\ref{fig:exp-anisotropy}d and present the results in Fig.~\ref{fig:exp-anisotropy}e, together with the $B_{\parallel}$-dependent $R_{xx}$ measured in the same device. The field magnitude corresponding to the peak of $\eta$ is close to the transport critical in-plane magnetic field ($B_{c,\parallel}$) observed in the resistivity measurement. Together with the systematic suppression of $\eta$ as superconductivity is further weakened by increasing the in-plane magnetic field beyond $B_{c,\parallel}$, this directly demonstrates that the observed two-fold anisotropy is intrinsically associated with superconductivity, consistent with a two-fold nematic pairing symmetry.

After identifying the two-fold anisotropy in $dI/dV(V=0)$ at different in-plane magnetic field magnitudes, we investigate the $B_{\parallel}$-dependence of $dI/dV(V=0)$ at various field directions $\theta_{B}$, as shown in Fig.~\ref{fig:exp-anisotropy}f. The two-fold anisotropy is most pronounced in the intermediate to high field range, consistent with the trend observed in Fig.~\ref{fig:exp-anisotropy}e. We plot $dI/dV(V=0)$ as a function of $B_{\parallel}$ at several representative $\theta_{B}$ in Fig.~\ref{fig:exp-anisotropy}g. For all field orientations, $dI/dV(V=0)$ initially increases slowly with the magnetic field, followed by an approximately linear dependence on $B_{\parallel}$, and eventually saturates at higher fields. We perform linear fits in the intermediate field range (dashed lines in Fig.~\ref{fig:exp-anisotropy}g), and extract the slope, $d^2I/(dVdB_{\parallel})(V=0)$, of the increasing trend as a function of $\theta_{B}$. Figure~\ref{fig:exp-anisotropy}h illustrates a prominent two-fold anisotropy with the same major axis direction as the $dI/dV(V=0)$ anisotropy (Fig.~\ref{fig:exp-anisotropy}b). Two additional weaker lobes, oriented approximately along the direction of $\SI{150}{^{\circ}}$ and $\SI{330}{^{\circ}}$, are also observed, although these features are absent in the anisotropic response of $dI/dV(V=0)$. A detailed understanding of these features is beyond the scope of the present work.


\begin{figure}
\includegraphics[width=0.8\textwidth]{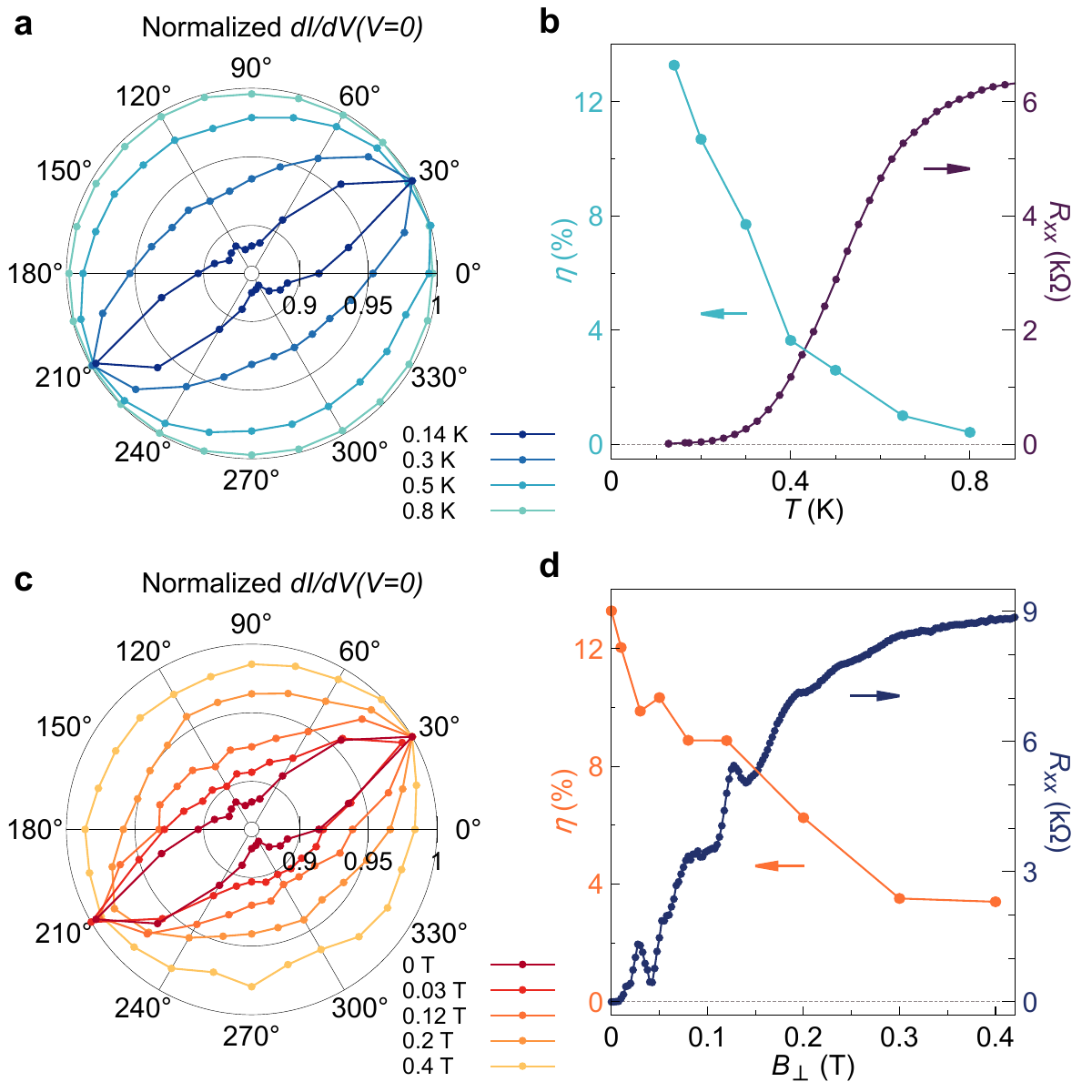}
\caption{\small Temperature and perpendicular magnetic field dependence of the tunneling conductance anisotropy. (a) Polar plot of the normalized angular-dependent $dI/dV(V=0)$ under various temperatures at $V_{BG}=\SI{-3.8}{\volt}$ and $B_{\parallel}=\SI{0.7}{\tesla}$. (b) Temperature dependence of the anisotropy ratio $\eta$ (light blue, left axis), extracted from (a), reveals that the anisotropy decreases continuously with increasing temperature, while $R_{xx}$ measured at $B_{\parallel}=\SI{0.7}{\tesla}$ is shown for comparison (dark purple, right axis). $\eta$ becomes nearly zero around the onset superconducting temperature $T_{onset}=\SI{0.84}{\kelvin}$, which marks the full suppression of superconductivity and recovery of the normal-state resistance. (c) Polar plot of the normalized angular-dependent $dI/dV(V=0)$ versus perpendicular magnetic field $B_{\perp}$ at $V_{BG}=\SI{-3.8}{\volt}$ and $B_{\parallel}=\SI{0.7}{\tesla}$. (d) $\eta$ versus $B_{\perp}$ (orange, left axis), extracted from (c), and $R_{xx}$ versus $B_{\perp}$ (dark blue, right axis). $\eta$ decreases gradually upon application of $B_{\perp}$. It maintains a relatively large anisotropy while MATTG is tuned out of the zero-resistance state, possibly owing to the existence of localized superconducting puddles. Notably, the peak feature in $\eta$ corresponds to the dip in $R_{xx}$ at $\sim\SI{50}{\milli\tesla}$. Similarly, the plateau in $\eta$ aligns with the plateau in $R_{xx}$ at $\sim\SI{100}{\milli\tesla}$ . This indicates that the fine features in the $B_{\perp}$ dependences of $\eta$ and $R_{xx}$ match closely.}
\label{fig:more-anisotropy}
\end{figure}

To gain further insight, we investigate the dependence of the normalized zero-bias tunneling conductance on the in-plane magnetic field angle $\theta_{B}$ under varying external parameters, such as temperature and perpendicular magnetic field. Figure~\ref{fig:more-anisotropy}a shows a polar plot of the normalized $dI/dV(V=0)$ as a function of $\theta_{B}$ at different temperatures (for fixed $B_{\parallel} = \SI{0.7}{\tesla}$ and $V_{BG} = \SI{-3.8}{\volt}$). We find that the magnitude of the anisotropy decreases sharply with increasing $T$ starting from the base temperature, followed by a slower suppression rate as $T$ increases further. To quantify this trend, we extract the temperature-dependent anisotropy ratio $\eta$ and compare it with the corresponding temperature dependence of $R_{xx}$ at $B_{\parallel}=\SI{0.7}{\tesla}$, as shown in Fig.~\ref{fig:more-anisotropy}b. The fact that $\eta$ drops rapidly as the temperature increases, while $R_{xx}$ is still in the zero-resistance state, indicates that the anisotropy is related to the low-energy nodal quasiparticles observed in the tunneling spectrum. Additionally, $\eta$ decreases continuously with increasing temperature and becomes nearly zero around the onset superconducting temperature $T_{onset}=\SI{0.84}{\kelvin}$, at which the transport resistance recovers to 95\% of its normal-state value (see Methods). While $\eta$ remains finite above the transport critical temperature $T_{c}=\SI{0.53}{\kelvin}$, this temperature range corresponds to the superconducting fluctuation regime rather than the normal state. The disappearance of $\eta$ at $T_{onset}$ demonstrates that the tunneling anisotropy vanishes only after superconductivity is fully suppressed, providing direct evidence that the observed anisotropy is intrinsic to the superconducting state and is not inherited from a normal-state nematic phase. Applying a perpendicular field $B_{\perp}$ is also expected to be detrimental to the superconducting phase coherence, thereby suppressing the tunneling conductance anisotropy. This is shown in the polar plot of the perpendicular magnetic field dependence of normalized $dI/dV(V=0)$ versus $\theta_{B}$ (Fig.~\ref{fig:more-anisotropy}c). As shown in Fig.~\ref{fig:more-anisotropy}d, which displays the anisotropy ratio $\eta$ as a function of $B_{\perp}$ overlaid with $R_{xx}$, $\eta$ decreases with $B_{\perp}$, consistent with the suppression of superconductivity in MATTG by a perpendicular field.

Having established the temperature and perpendicular magnetic field dependence of the tunneling conductance anisotropy, we now examine its evolution with carrier density across the superconducting dome, which could offer further insight into the nature of superconductivity in MATTG. Figure~\ref{fig:anisotropy-across-dome}a presents the 2D map of $R_{xx}$ as a function of $V_{BG}$ and $T$, centered around the hole-doped superconducting dome. The color ticks along the bottom axis indicate the gate voltages corresponding to the measurements in Fig.~\ref{fig:anisotropy-across-dome}b–q, which were performed at an elevated temperature of $T=\SI{300}{\milli\kelvin}$, instead of the base temperature used in Fig.~\ref{fig:exp-anisotropy}, to suppress disorder-induced sharp features around zero-bias and enable a consistent comparison across gate voltages. Note that the previous tunneling conductance anisotropy measurements reported in this study were performed on the overdoped side of the superconducting dome, at $V_{BG}=\SI{-3.8}{V}$, where the major axis of the anisotropy is oriented along $\SI{30}{^{\circ}}-\SI{210}{^{\circ}}$ (Fig.~\ref{fig:anisotropy-across-dome}d). We define the direction of the major axis $\theta_{N}$ as the first angle in the above expression, $\theta_{N}=30{^\circ}$ at $V_{BG}=\SI{-3.8}{V}$, constrained to lie within the range $0{^\circ}-180{^\circ}$ for any doping. $\theta_{N}$ is obtained by fitting the tunneling data with an ellipse function (see Methods; fitted ellipse functions shown as the gray dashed lines in Fig.~\ref{fig:anisotropy-across-dome}b-q). As MATTG enters the superconducting state from the overdoped side, an anisotropic response in the tunneling conductance progressively develops, with the anisotropy ratio increasing from $\eta\sim\SI{1}{\%}$ at $V_{BG}=\SI{-4.1}{V}$ (Fig.~\ref{fig:anisotropy-across-dome}b) to $\eta\sim\SI{7}{\%}$ at $V_{BG}=\SI{-3.8}{V}$ (Fig.~\ref{fig:anisotropy-across-dome}d). Further doping into the superconducting dome, $\theta_{N}$ gradually rotates to $\sim{55}{^{\circ}}$ at $V_{BG}=\SI{-3.5}{V}$ (Fig.~\ref{fig:anisotropy-across-dome}f), and undergoes an abrupt shift to $\sim{135}{^{\circ}}$ at $V_{BG}=\SI{-3.475}{V}$ (Fig.~\ref{fig:anisotropy-across-dome}g) accompanied by a peak in $\eta\sim\SI{11}{\%}$. As the system approaches the optimal doping, $\theta_{N}$ begins to fluctuate around $\SI{90}{^{\circ}}$, as observed in Fig.~\ref{fig:anisotropy-across-dome}h–l. Eventually, $\theta_{N}$ returns to $\sim\SI{50}{^{\circ}}$ at $V_{BG}=\SI{-3.1}{V}$ (Fig.~\ref{fig:anisotropy-across-dome}m), and remains near this orientation in the underdoped regime. At the same time, the anisotropy ratio gradually decreases as the system exits the superconducting state. These observations indicate that the major axis of the tunneling conductance anisotropy is not confined to a specific direction, but rather rotates across a wide angular range as doping changes. The abrupt rotation by nearly $\SI{90}{^{\circ}}$ at around $V_{BG} = \SI{-3.5}{V}$ is particularly striking, suggesting that the origin of the anisotropy is intrinsic and cannot be simply attributed to residual strain in the MATTG device. A similar rotation of a two-fold anisotropic response across the superconducting dome was observed in MATBG\cite{cao_nematicity_2021}, but in that case, the response was the in-plane critical field through transport measurements, which is not directly sensitive to the low-energy excitations. We note that a doping-induced rotation of the nematic director was also observed in cuprate thin films via resistivity anisotropy measurements, but in the normal state \cite{Wu2017}. Here, we observe nematic order only inside the superconducting state.

To further examine the evolution of the anisotropy ratio and the major axis across the superconducting dome, we extract $\eta$ and $\theta_{N}$ at each carrier density (Fig.~\ref{fig:anisotropy-across-dome}r). The anisotropy ratios, $\eta_{1}$ and $\eta_{2}$, are obtained from measurements using two distinct contact pairs, with the color opacity representing the strength of the anisotropy. The similar $V_{BG}$ dependence of $\theta_{N}$ corresponding to $\eta_{1}$ and $\eta_{2}$ reveals that the overall trend of the major axis direction is independent of the choice of contact pairs used in the tunneling measurements, providing further evidence that the observed anisotropy is not a result of strain. Thus, we conclude that the doping-dependent tunneling conductance anisotropy reflects a nematic superconducting ground state with two-fold pairing symmetry, whose preferred major axis is strongly influenced by both intrinsic factors, such as the doping-dependent Fermi surface geometry and electronic correlations, and extrinsic factors, including strain introduced during the nanofabrication process\cite{chakraborty2023strain}. We note that some strain is likely present in the sample, otherwise domains with equal population would be present, resulting in a net six-fold symmetric response. While heterostrain is known to have a strong impact on the DOS of twisted graphene devices \cite{Wang2023}, we can rule it out from being the source of the anisotropy observed here. If strain was the driver of the observed two-fold anisotropy, then one would not expect the sudden changes in the anisotropy direction shown in Fig.~\ref{fig:anisotropy-across-dome}r, nor the independence of the anisotropy ratio and anisotropy direction on the choice of contacts. We can also rule out device inhomogeneity as the source of the anisotropy, since the observed anisotropy increases as temperature is lowered. If the anisotropy was due to an inhomogeneous distribution of regions with different local $T_c$ values, the anisotropy would be expected to be largest close to $T_c$\cite{Vaglio1993}.

We now discuss the implications of our observations for the pairing state of MATTG. Clearly, the intrinsic origin of the two-fold anisotropy of the zero-bias tunneling conductance requires a nematic pairing state. Moreover, the main contributions to the zero-bias conductance deep inside the superconducting state generally come from nodal quasiparticle excitations. Consequently, our experimental results imply a nodal nematic pairing state. Since this state spontaneously breaks the sixfold rotational symmetry of the system, the positions of the nodes shift away from the high-symmetry positions. To further support this assessment, and to disentangle between different types of nematic states, we perform a phenomenological analysis of the zero-bias tunneling conductance. Some features of the tunneling conductance can directly reflect general properties of the superconductor, such as the shape of the $dI/dV$ curve at small bias, which provides information about the presence of gap nodes. However, the finite $dI/dV$ value at zero bias is not a universal property of the superconducting state, but a result of junction-specific pair-breaking processes. Theoretically, the latter is often described by the phenomenological Dynes model\cite{Dynes1984,dynes_direct_1978}. In order to draw general conclusions from our measurements without resorting to specific pairing mechanisms or gap functions, which can assume many forms in MATTG\cite{Scheurer2020, Khalaf2022, lake_pairing_2022}, here we focus on the broader question of which pairing symmetries give rise to a two-fold anisotropic $dI/dV(V=0)$ as a function of a rotating in-plane field $\vb{B}_{\parallel}$. To accomplish this, we assume a Dynes model with uniform pair-breaking parameter and a generic isotropic Fermi surface (see Methods for details). In this case, a two-fold anisotropy cannot be generated unless the gap function itself is anisotropic, i.e., nematic. However, as we explain now, a nematic pairing state is by itself not enough to ensure a sizable two-fold-anisotropic $dI/dV(V=0)$ as a function of $\vb{B}_{\parallel}$.

To proceed, we need to establish how the magnetic field couples to the superconducting quasiparticles, which at low temperatures and low energies come from the nodal regions of the Fermi surface. Generally, there is an orbital coupling, giving rise to a Doppler shift in the energies of the quasiparticles\cite{volovik_superconductivity_1993} and Zeeman coupling through momentum boost. Although the former is often the dominant mechanism, for in-plane magnetic fields in atomically thin superconductors such as MATTG, this effect is not present because there is no room for vortices orthogonal to the plane. This leaves the Zeeman coupling as the only possibility. In twisted graphene systems, the Zeeman coupling can arise from spin, as in bulk triplet superconductors, or from valley (`orbital') magnetization, as observed in MATBG\cite{cao_nematicity_2021}. However, in MATTG, mirror symmetry suppresses the pair-breaking effect contributed by orbital-Zeeman coupling. Therefore, for our phenomenological analysis, we focus only on the spin-Zeeman coupling, as it remains non-zero even in the ideal regime of mirror-symmetric MATTG.

In this case, in order for the Bogoliubov quasiparticle dispersion to be sensitive to an in-plane magnetic field, the pairing state needs to be triplet or at least have a sizable spin-triplet component, since the in-plane field direction only appears in the quasiparticle dispersion through its coupling to the in-plane component of the d-vector $\vb{d}(\vb{k})$ via $\lvert{\vb{d}_{\parallel} \vdot \vb{B}_{\parallel}}\rvert^2$ (see Methods). This expression has important consequences. First, if SOC is absent and $\vb{d}_{\parallel}$ is free to point along any direction in spin space, in the presence of a field it will inevitably be locked to the direction perpendicular to $\vb{B}_{\parallel}$. Consequently, the tunneling conductance at zero bias has no anisotropy. Therefore, the observed anisotropy requires a small but non-zero SOC to lock the d-vector to the lattice, as anticipated in the discussion above. Moreover, the d-vector must have an in-plane component, which itself must be nematic and break six-fold rotation symmetry, in order for a response function to display two-fold anisotropy.

\begin{figure}
\includegraphics[width=1\textwidth]{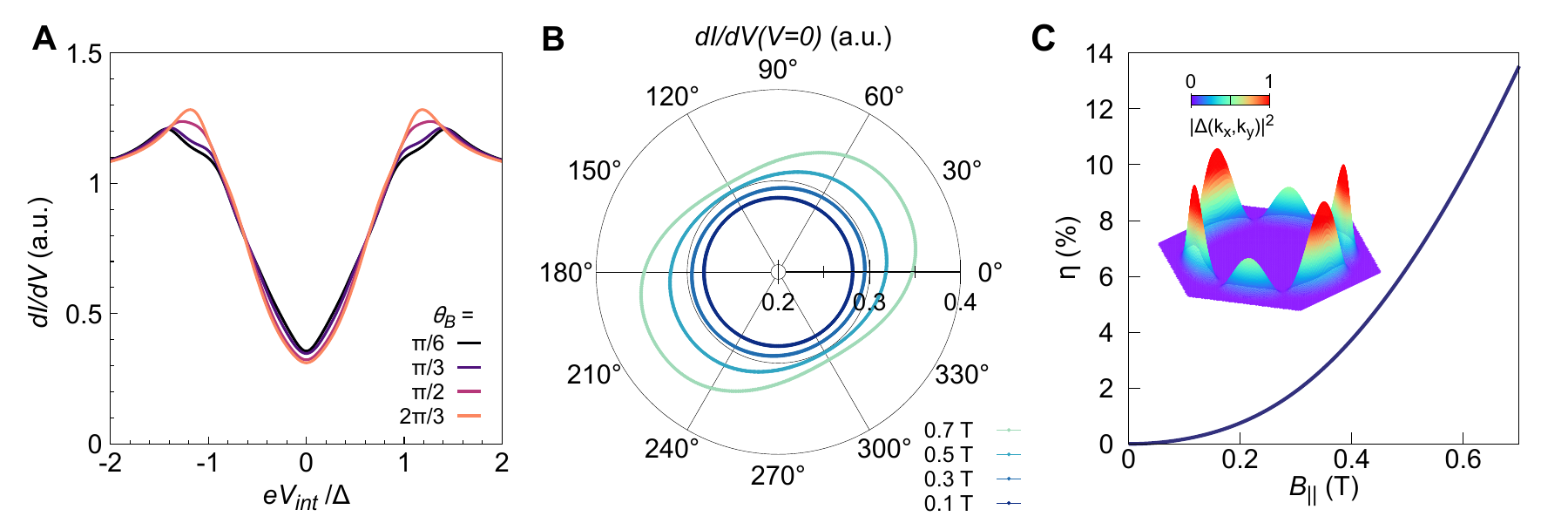}
\caption{\small Theoretical tunneling conductance of a nematic triplet superconducting state whose d-vector is in-plane. (a) $dI/dV$ as a function of $eV_{int}$, in units of the superconducting gap $\Delta = \SI{0.15}{meV}$ for various angles $\theta_{B}$ when $B_{\parallel} = \SI{0.7}{\tesla}$. The main changes in $dI/dV$ are near zero bias and near the coherence peaks. (b) Polar plot of $dI/dV(V=0)$ versus $\theta_{B}$, with $B_{\parallel}$ ranging from $\SI{0.1}{\tesla}$ to $\SI{0.7}{\tesla}$. A two-fold anisotropy is observed, consistent with the experimental observation in Fig.~\ref{fig:exp-anisotropy}b. (c) Anisotropy ratio $\eta$ as a function of the $B_{\parallel}$ magnitude, while keeping the superconducting gap magnitude constant. Inset: the spin-triplet superconducting gap function $|\Delta(k_x,k_y)|^2$. All parameters are given in the Methods section.}
\label{fig:tunneling-DOS}
\end{figure}

These symmetry-based constraints rule out singlet states, non-nematic triplet states, and nematic triplet states with out-of-plane d-vectors, leaving nematic triplet states with in-plane $\vb{d}_{\parallel}$ as the main candidate consistent with our observation. From a group-theory perspective, such a state transforms as the irreducible representation $E_{2u}$ of the hexagonal point group $D_{6h}$. The precise form of the gap function, however, is not entirely enforced by symmetry and requires additional information about the Fermi surface and the microscopic interactions involved. Nevertheless, to gain further insight, it is instructive to consider the simplest case of a featureless Fermi surface with six-fold anisotropy. In this case, depending on parameters, the gap function can be fully gapped or nodal. Our calculation of the zero-bias tunneling conductance shows that the two-fold anisotropy is negligible for a fully gapped state with a nearly isotropic gap, but sizable for a nodal state or a quasi-nodal state in which the gap is strongly anisotropic with very small gap minima (see Methods and Extended Data Fig.~\ref*{fig:Full-gap-no-nema}). This is not unexpected, since the zero-bias tunneling conductance primarily arises from pair-breaking processes involving the low-energy quasiparticle excitations, which are strongly suppressed for fully gapped pairing states. The requirement of nodes or quasi-nodes is consistent with the V-shaped tunneling conductance observed, when broadening is also included. To illustrate these results, in Fig.~\ref{fig:tunneling-DOS} we show a model calculation for the $dI/dV$ spectrum of a nematic triplet superconductor with an in-plane d-vector in the presence of an in-plane magnetic field (see Methods for details). Fig.~\ref{fig:tunneling-DOS}a shows that the field direction mostly impacts the regions of the spectrum that are near zero bias and near the coherence peaks. The polar plot of the zero-bias tunneling conductance, displayed in Fig.~\ref{fig:tunneling-DOS}b, shows the characteristic two-fold anisotropy observed in our experiments as in Fig.~\ref{fig:exp-anisotropy}b. Moreover, the magnitude of the anisotropy $\eta$ increases with the magnetic field, as shown in Fig.~\ref{fig:tunneling-DOS}c, since more low-energy quasiparticles undergo pair-breaking when the field is larger. This gradual increase of $\eta$ in the low-field regime is qualitatively consistent with the experimental observations in Fig.~\ref{fig:exp-anisotropy}e. The present phenomenological model is intended only to describe the low-field response of the tunneling conductance, as the microscopic pair-breaking mechanism of an in-plane magnetic field in MATTG remains unclear. The suppression of the anisotropy ratio at higher fields can also be qualitatively captured if we include a phenomenological field suppression of the superconducting gap\cite{douglass1961magnetic,bardeen1962critical}, as shown in Extended Data Fig.~\ref*{fig:Parameter-tuning}, with a parameter-dependent peak position of $\eta$.

The observed two-fold anisotropy of the zero-bias tunneling conductance under an in-plane magnetic field provides strong evidence for a nodal, nematic pairing state in MATTG.
Provided that mirror symmetry-breaking is negligible and SOC is finite, our phenomenological analysis further implies that this pairing state features a triplet component with an in-plane d-vector. We note that there may be various pairing functions that may satisfy these conditions, including superconducting states that mix singlet and triplet components. Moreover, while the analysis performed here focused on symmetry and on a generic Fermi surface, the electronic degrees of freedom present in twisted graphene systems may enable other types of pairing functions with nodes, nematicity, and triplet pairing\cite{liu2025nodal}. Finally, if the mirror symmetry-breaking is sufficiently strong in the device, the orbital-Zeeman effect may induce a significant depairing energy, in which case non-triplet pairing states could also become sensitive to in-plane fields. Notwithstanding these caveats, the constraint of an intrinsic nodal nematic pairing state imposed by our experiments strongly narrows down the possible gap functions, which in turn further constrains the possible underlying pairing mechanisms. More broadly, the experimental platform developed in this work provides a powerful method to simultaneously probe symmetry-breaking superconducting states and low-energy superconducting quasiparticles in 2D materials. For future studies, phase-sensitive experiments capable of detecting directional phase shifts in the superconducting gap function would offer further insights into the exact form of the pairing symmetry in moiré graphene.

\bibliographystyle{Nature}
\bibliography{reference_corrected}

@article{hamill2021two,
  title={Two-fold symmetric superconductivity in few-layer NbSe2},
  author={Hamill, Alex and Heischmidt, Brett and Sohn, Egon and Shaffer, Daniel and Tsai, Kan-Ting and Zhang, Xi and Xi, Xiaoxiang and Suslov, Alexey and Berger, Helmuth and Forr{\'o}, L{\'a}szl{\'o} and others},
  journal={Nature physics},
  volume={17},
  number={8},
  pages={949--954},
  year={2021},
  publisher={Nature Publishing Group UK London}
}

@article{li2017nematic,
  title={Nematic superconducting state in iron pnictide superconductors},
  author={Li, Jun and Pereira, Paulo J and Yuan, Jie and Lv, Yang-Yang and Jiang, Mei-Ping and Lu, Dachuan and Lin, Zi-Quan and Liu, Yong-Jie and Wang, Jun-Feng and Li, Liang and others},
  journal={Nature communications},
  volume={8},
  number={1},
  pages={1880},
  year={2017},
  publisher={Nature Publishing Group UK London}
}

@article{douglass1961magnetic,
  title={Magnetic field dependence of the superconducting energy gap},
  author={Douglass Jr, DH},
  journal={Physical Review Letters},
  volume={6},
  number={7},
  pages={346},
  year={1961},
  publisher={APS}
}

@article{bardeen1962critical,
  title={Critical fields and currents in superconductors},
  author={Bardeen, John},
  journal={Reviews of modern physics},
  volume={34},
  number={4},
  pages={667},
  year={1962},
  publisher={APS}
}

@article{chakraborty2023strain,
  title={Strain-tuned quantum criticality in electronic Potts-nematic systems},
  author={Chakraborty, Anzumaan R and Fernandes, Rafael M},
  journal={Physical Review B},
  volume={107},
  number={19},
  pages={195136},
  year={2023},
  publisher={APS}
}

@article{xie2019spectroscopic,
  title={Spectroscopic signatures of many-body correlations in magic-angle twisted bilayer graphene},
  author={Xie, Yonglong and Lian, Biao and J{\"a}ck, Berthold and Liu, Xiaomeng and Chiu, Cheng-Li and Watanabe, Kenji and Taniguchi, Takashi and Bernevig, B Andrei and Yazdani, Ali},
  journal={Nature},
  volume={572},
  number={7767},
  pages={101--105},
  year={2019},
  publisher={Nature Publishing Group UK London}
}

@article{palstra1988angular,
  title={Angular dependence of the upper critical field of {Bi}$_{\textrm{2.2}}${Sr}$_{\textrm{2}}${Ca}$_{\textrm{0.8}}${Cu}$_{\textrm{2}}${O}$_{\textrm{8+$\delta$}}$},
  author={Palstra, TTM and Batlogg, Bertram and Schneemeyer, LF and Van Dover, RB and Waszczak, Joseph V},
  journal={Physical Review B},
  volume={38},
  number={7},
  pages={5102},
  year={1988},
  publisher={APS}
}

@book{vollhardt2013superfluid,
  title={The Superfluid Phases of Helium 3},
  author={Vollhardt, Dieter and Wolfle, Peter},
  year={2013},
  publisher={Courier Corporation}
}

@article{moler1994magnetic,
  title={Magnetic field dependence of the density of states of {Y}{Ba}$_{\textrm{2}}${Cu}$_{\textrm{3}}${O}$_{\textrm{6.95}}$ as determined from the specific heat},
  author={Moler, Kathryn A and Baar, DJ and Urbach, JS and Liang, Ruixing and Hardy, WN and Kapitulnik, A},
  journal={Physical review letters},
  volume={73},
  number={20},
  pages={2744},
  year={1994},
  publisher={APS}
}

@article{Vojta2000,
  title = {Quantum Phase Transitions in {{$\mathit{d}$}}-Wave Superconductors},
  author = {Vojta, Matthias and Zhang, Ying and Sachdev, Subir},
  journal = {Phys. Rev. Lett.},
  volume = {85},
  issue = {23},
  pages = {4940--4943},
  numpages = {0},
  year = {2000},
  month = {Dec},
  publisher = {American Physical Society},
  doi = {10.1103/PhysRevLett.85.4940},
  url = {https://link.aps.org/doi/10.1103/PhysRevLett.85.4940}
}

@article{Kim2008,
  title = {Theory of the nodal nematic quantum phase transition in superconductors},
  author = {Kim, Eun-Ah and Lawler, Michael J. and Oreto, Paul and Sachdev, Subir and Fradkin, Eduardo and Kivelson, Steven A.},
  journal = {Phys. Rev. B},
  volume = {77},
  issue = {18},
  pages = {184514},
  numpages = {9},
  year = {2008},
  month = {May},
  publisher = {American Physical Society},
  doi = {10.1103/PhysRevB.77.184514},
  url = {https://link.aps.org/doi/10.1103/PhysRevB.77.184514}
}

@article{Yu_Sarma2023,
  title = {Euler-obstructed nematic nodal superconductivity in twisted bilayer graphene},
  author = {Yu, Jiabin and Xie, Ming and Wu, Fengcheng and Das Sarma, Sankar},
  journal = {Phys. Rev. B},
  volume = {107},
  issue = {20},
  pages = {L201106},
  numpages = {7},
  year = {2023},
  month = {May},
  publisher = {American Physical Society},
  doi = {10.1103/PhysRevB.107.L201106},
  url = {https://link.aps.org/doi/10.1103/PhysRevB.107.L201106}
}

@article{Song2024,
  title = {Molecular Pairing in Twisted Bilayer Graphene Superconductivity},
  author = {Wang, Yi-Jie and Zhou, Geng-Dong and Peng, Shi-Yu and Lian, Biao and Song, Zhi-Da},
  journal = {Phys. Rev. Lett.},
  volume = {133},
  issue = {14},
  pages = {146001},
  numpages = {9},
  year = {2024},
  month = {Sep},
  publisher = {American Physical Society},
  doi = {10.1103/PhysRevLett.133.146001},
  url = {https://link.aps.org/doi/10.1103/PhysRevLett.133.146001}
}

@article{Wang_Fernandes2021,
  title = {Topological and nematic superconductivity mediated by ferro-{SU}(4) fluctuations in twisted bilayer graphene},
  author = {Wang, Yuxuan and Kang, Jian and Fernandes, Rafael M.},
  journal = {Phys. Rev. B},
  volume = {103},
  issue = {2},
  pages = {024506},
  numpages = {15},
  year = {2021},
  month = {Jan},
  publisher = {American Physical Society},
  doi = {10.1103/PhysRevB.103.024506},
  url = {https://link.aps.org/doi/10.1103/PhysRevB.103.024506}
}

@article{keimer2015quantum,
  title={From quantum matter to high-temperature superconductivity in copper oxides},
  author={Keimer, Bernhard and Kivelson, Steven A and Norman, Michael R and Uchida, Shinichi and Zaanen, J},
  journal={Nature},
  volume={518},
  number={7538},
  pages={179--186},
  year={2015},
  publisher={Nature Publishing Group UK London}
}

@article{zhang2025angular,
  title={Angular Interplay of Nematicity, Superconductivity, and Strange Metallicity in a Moiré Flat Band},
  author={Zhang, Naiyuan J and Nosov, Pavel A and Sommer, Ophelia Evelyn and Wang, Yibang and Watanabe, Kenji and Taniguchi, Takashi and Khalaf, Eslam and Li, JIA},
  journal={arXiv preprint arXiv:2503.15767},
  year={2025}
}

@article{liu2025nodal,
  title={Nodal nematic superconductivity in multiple flat-band systems},
  author={Liu, Chao-Xing and Bernevig, B Andrei},
  journal={Physical Review B},
  volume={111},
  number={2},
  pages={L020502},
  year={2025},
  publisher={APS}
}

@article{andrei_marvels_2021,
	title = {The marvels of moiré materials},
	volume = {6},
	copyright = {2021 Springer Nature Limited},
	issn = {2058-8437},
	url = {https://www.nature.com/articles/s41578-021-00284-1},
	doi = {10.1038/s41578-021-00284-1},
	language = {en},
	number = {3},
	urldate = {2024-10-09},
	journal = {Nature Reviews Materials},
	author = {Andrei, Eva Y. and Efetov, Dmitri K. and Jarillo-Herrero, Pablo and MacDonald, Allan H. and Mak, Kin Fai and Senthil, T. and Tutuc, Emanuel and Yazdani, Ali and Young, Andrea F.},
	month = mar,
	year = {2021},
	note = {Publisher: Nature Publishing Group},
	pages = {201--206},
}

@article{cao_unconventional_2018,
	title = {Unconventional superconductivity in magic-angle graphene superlattices},
	volume = {556},
	copyright = {2018 Macmillan Publishers Limited, part of Springer Nature. All rights reserved.},
	issn = {1476-4687},
	url = {https://www.nature.com/articles/nature26160},
	doi = {10.1038/nature26160},
	language = {en},
	number = {7699},
	urldate = {2024-12-23},
	journal = {Nature},
	author = {Cao, Yuan and Fatemi, Valla and Fang, Shiang and Watanabe, Kenji and Taniguchi, Takashi and Kaxiras, Efthimios and Jarillo-Herrero, Pablo},
	month = apr,
	year = {2018},
	note = {Publisher: Nature Publishing Group},
	pages = {43--50},
}

@article{park_tunable_2021,
	title = {Tunable strongly coupled superconductivity in magic-angle twisted trilayer graphene},
	volume = {590},
	copyright = {2021 The Author(s), under exclusive licence to Springer Nature Limited},
	issn = {1476-4687},
	url = {https://www.nature.com/articles/s41586-021-03192-0},
	doi = {10.1038/s41586-021-03192-0},
	language = {en},
	number = {7845},
	urldate = {2024-12-23},
	journal = {Nature},
	author = {Park, Jeong Min and Cao, Yuan and Watanabe, Kenji and Taniguchi, Takashi and Jarillo-Herrero, Pablo},
	month = feb,
	year = {2021},
	note = {Publisher: Nature Publishing Group},
	pages = {249--255},
}

@article{hao_electric_2021,
	title = {Electric field–tunable superconductivity in alternating-twist magic-angle trilayer graphene},
	volume = {371},
	url = {https://www.science.org/doi/full/10.1126/science.abg0399},
	doi = {10.1126/science.abg0399},
	number = {6534},
	urldate = {2024-12-23},
	journal = {Science},
	author = {Hao, Zeyu and Zimmerman, A. M. and Ledwith, Patrick and Khalaf, Eslam and Najafabadi, Danial Haie and Watanabe, Kenji and Taniguchi, Takashi and Vishwanath, Ashvin and Kim, Philip},
	month = mar,
	year = {2021},
	note = {Publisher: American Association for the Advancement of Science},
	pages = {1133--1138},
}

@article{park_robust_2022,
	title = {Robust superconductivity in magic-angle multilayer graphene family},
	volume = {21},
	copyright = {2022 The Author(s), under exclusive licence to Springer Nature Limited},
	issn = {1476-4660},
	url = {https://www.nature.com/articles/s41563-022-01287-1},
	doi = {10.1038/s41563-022-01287-1},
	language = {en},
	number = {8},
	urldate = {2024-12-23},
	journal = {Nature Materials},
	author = {Park, Jeong Min and Cao, Yuan and Xia, Li-Qiao and Sun, Shuwen and Watanabe, Kenji and Taniguchi, Takashi and Jarillo-Herrero, Pablo},
	month = aug,
	year = {2022},
	note = {Publisher: Nature Publishing Group},
	pages = {877--883},
}

@article{zhang_promotion_2022,
	title = {Promotion of superconductivity in magic-angle graphene multilayers},
	volume = {377},
	url = {https://www.science.org/doi/full/10.1126/science.abn8585},
	doi = {10.1126/science.abn8585},
	number = {6614},
	urldate = {2024-12-23},
	journal = {Science},
	author = {Zhang, Yiran and Polski, Robert and Lewandowski, Cyprian and Thomson, Alex and Peng, Yang and Choi, Youngjoon and Kim, Hyunjin and Watanabe, Kenji and Taniguchi, Takashi and Alicea, Jason and von Oppen, Felix and Refael, Gil and Nadj-Perge, Stevan},
	month = sep,
	year = {2022},
	note = {Publisher: American Association for the Advancement of Science},
	pages = {1538--1543},
}

@article{oh_evidence_2021,
	title = {Evidence for unconventional superconductivity in twisted bilayer graphene},
	volume = {600},
	copyright = {2021 The Author(s), under exclusive licence to Springer Nature Limited},
	issn = {1476-4687},
	url = {https://www.nature.com/articles/s41586-021-04121-x},
	doi = {10.1038/s41586-021-04121-x},
	language = {en},
	number = {7888},
	urldate = {2024-12-23},
	journal = {Nature},
	author = {Oh, Myungchul and Nuckolls, Kevin P. and Wong, Dillon and Lee, Ryan L. and Liu, Xiaomeng and Watanabe, Kenji and Taniguchi, Takashi and Yazdani, Ali},
	month = dec,
	year = {2021},
	note = {Publisher: Nature Publishing Group},
	pages = {240--245},
}

@article{kim_evidence_2022,
	title = {Evidence for unconventional superconductivity in twisted trilayer graphene},
	volume = {606},
	copyright = {2022 The Author(s), under exclusive licence to Springer Nature Limited},
	issn = {1476-4687},
	url = {https://www.nature.com/articles/s41586-022-04715-z},
	doi = {10.1038/s41586-022-04715-z},
	language = {en},
	number = {7914},
	urldate = {2024-12-23},
	journal = {Nature},
	author = {Kim, Hyunjin and Choi, Youngjoon and Lewandowski, Cyprian and Thomson, Alex and Zhang, Yiran and Polski, Robert and Watanabe, Kenji and Taniguchi, Takashi and Alicea, Jason and Nadj-Perge, Stevan},
	month = jun,
	year = {2022},
	note = {Publisher: Nature Publishing Group},
	pages = {494--500},
}

@article{cao_pauli-limit_2021,
	title = {Pauli-limit violation and re-entrant superconductivity in moiré graphene},
	volume = {595},
	copyright = {2021 The Author(s), under exclusive licence to Springer Nature Limited},
	issn = {1476-4687},
	url = {https://www.nature.com/articles/s41586-021-03685-y},
	doi = {10.1038/s41586-021-03685-y},
	language = {en},
	number = {7868},
	urldate = {2024-12-23},
	journal = {Nature},
	author = {Cao, Yuan and Park, Jeong Min and Watanabe, Kenji and Taniguchi, Takashi and Jarillo-Herrero, Pablo},
	month = jul,
	year = {2021},
	note = {Publisher: Nature Publishing Group},
	pages = {526--531},
}

@article{cao_nematicity_2021,
	title = {Nematicity and competing orders in superconducting magic-angle graphene},
	volume = {372},
	url = {https://www.science.org/doi/full/10.1126/science.abc2836},
	doi = {10.1126/science.abc2836},
	number = {6539},
	urldate = {2024-12-23},
	journal = {Science},
	author = {Cao, Yuan and Rodan-Legrain, Daniel and Park, Jeong Min and Yuan, Noah F. Q. and Watanabe, Kenji and Taniguchi, Takashi and Fernandes, Rafael M. and Fu, Liang and Jarillo-Herrero, Pablo},
	month = apr,
	year = {2021},
	note = {Publisher: American Association for the Advancement of Science},
	pages = {264--271},
}

@article{dynes_direct_1978,
	title = {Direct {Measurement} of {Quasiparticle}-{Lifetime} {Broadening} in a {Strong}-{Coupled} {Superconductor}},
	volume = {41},
	url = {https://link.aps.org/doi/10.1103/PhysRevLett.41.1509},
	doi = {10.1103/PhysRevLett.41.1509},
	number = {21},
	urldate = {2024-12-23},
	journal = {Physical Review Letters},
	author = {Dynes, R. C. and Narayanamurti, V. and Garno, J. P.},
	month = nov,
	year = {1978},
	note = {Publisher: American Physical Society},
	pages = {1509--1512},
}

@article{volovik_superconductivity_1993,
author = {Volovik, G. E.},
title = {Superconductivity with lines of gap nodes: density of states in the vortex},
journal = {JETP Letters},
volume = {58},
number = {6},
pages = {457-461},
year = {1993},
URL = {http://jetpletters.ru/ps/1189/article\_17954.shtml}
}

@article{tanaka_superfluid_2025,
	title = {Superfluid stiffness of magic-angle twisted bilayer graphene},
	volume = {638},
	copyright = {2025 The Author(s), under exclusive licence to Springer Nature Limited},
	issn = {1476-4687},
	url = {https://www.nature.com/articles/s41586-024-08494-7},
	doi = {10.1038/s41586-024-08494-7},
	language = {en},
	number = {8049},
	urldate = {2025-03-20},
	journal = {Nature},
	author = {Tanaka, Miuko and Wang, Joel Î-j and Dinh, Thao H. and Rodan-Legrain, Daniel and Zaman, Sameia and Hays, Max and Almanakly, Aziza and Kannan, Bharath and Kim, David K. and Niedzielski, Bethany M. and Serniak, Kyle and Schwartz, Mollie E. and Watanabe, Kenji and Taniguchi, Takashi and Orlando, Terry P. and Gustavsson, Simon and Grover, Jeffrey A. and Jarillo-Herrero, Pablo and Oliver, William D.},
	month = feb,
	year = {2025},
	note = {Publisher: Nature Publishing Group},
	pages = {99--105},
}

@article{banerjee_superfluid_2025,
	title = {Superfluid stiffness of twisted trilayer graphene superconductors},
	volume = {638},
	copyright = {2025 The Author(s), under exclusive licence to Springer Nature Limited},
	issn = {1476-4687},
	url = {https://www.nature.com/articles/s41586-024-08444-3},
	doi = {10.1038/s41586-024-08444-3},
	language = {en},
	number = {8049},
	urldate = {2025-03-20},
	journal = {Nature},
	author = {Banerjee, Abhishek and Hao, Zeyu and Kreidel, Mary and Ledwith, Patrick and Phinney, Isabelle and Park, Jeong Min and Zimmerman, Andrew and Wesson, Marie E. and Watanabe, Kenji and Taniguchi, Takashi and Westervelt, Robert M. and Yacoby, Amir and Jarillo-Herrero, Pablo and Volkov, Pavel A. and Vishwanath, Ashvin and Fong, Kin Chung and Kim, Philip},
	month = feb,
	year = {2025},
	note = {Publisher: Nature Publishing Group},
	pages = {93--98},
}

@article{park_simultaneous_2025,
  title={Experimental evidence for nodal superconducting gap in moir{\'e} graphene},
  author={Park, Jeong Min and Sun, Shuwen and Watanabe, Kenji and Taniguchi, Takashi and Jarillo-Herrero, Pablo},
  journal={Science},
  volume={391},
  number={6780},
  pages={79--83},
  year={2026},
  publisher={American Association for the Advancement of Science}
}

@article{kim2026resolving,
  title={Resolving intervalley gaps and many-body resonances in moir{\'e} superconductors},
  author={Kim, Hyunjin and Rai, Gautam and Crippa, Lorenzo and C{\u{a}}lug{\u{a}}ru, Dumitru and Hu, Haoyu and Choi, Youngjoon and Kong, Lingyuan and Baum, Eli and Zhang, Yiran and Holleis, Ludwig and others},
  journal={Nature},
  volume={650},
  pages={592–598},
  year={2026},
  publisher={Nature Publishing Group UK London}
}

@article{yonezawa_thermodynamic_2017,
	title = {Thermodynamic evidence for nematic superconductivity in {Cu}$_{\textrm{x}}${Bi}$_{\textrm{2}}${Se}$_{\textrm{3}}$},
	volume = {13},
	copyright = {2016 Springer Nature Limited},
	issn = {1745-2481},
	url = {https://www.nature.com/articles/nphys3907},
	doi = {10.1038/nphys3907},
	language = {en},
	number = {2},
	urldate = {2025-08-13},
	journal = {Nature Physics},
	author = {Yonezawa, Shingo and Tajiri, Kengo and Nakata, Suguru and Nagai, Yuki and Wang, Zhiwei and Segawa, Kouji and Ando, Yoichi and Maeno, Yoshiteru},
	month = feb,
	year = {2017},
	note = {Publisher: Nature Publishing Group},
	pages = {123--126},
}

@article{matano_spin-rotation_2016,
	title = {Spin-rotation symmetry breaking in the superconducting state of {Cu}$_{\textrm{x}}${Bi}$_{\textrm{2}}${Se}$_{\textrm{3}}$},
	volume = {12},
	copyright = {2016 Springer Nature Limited},
	issn = {1745-2481},
	url = {https://www.nature.com/articles/nphys3781},
	doi = {10.1038/nphys3781},
	language = {en},
	number = {9},
	urldate = {2025-08-13},
	journal = {Nature Physics},
	author = {Matano, K. and Kriener, M. and Segawa, K. and Ando, Y. and Zheng, Guo-qing},
	month = sep,
	year = {2016},
	note = {Publisher: Nature Publishing Group},
	pages = {852--854},
}

@article{cao_correlated_2018,
	title = {Correlated insulator behaviour at half-filling in magic-angle graphene superlattices},
	volume = {556},
	copyright = {2018 Macmillan Publishers Limited, part of Springer Nature. All rights reserved.},
	issn = {1476-4687},
	url = {https://www.nature.com/articles/nature26154},
	doi = {10.1038/nature26154},
	language = {en},
	number = {7699},
	urldate = {2025-08-22},
	journal = {Nature},
	author = {Cao, Yuan and Fatemi, Valla and Demir, Ahmet and Fang, Shiang and Tomarken, Spencer L. and Luo, Jason Y. and Sanchez-Yamagishi, Javier D. and Watanabe, Kenji and Taniguchi, Takashi and Kaxiras, Efthimios and Ashoori, Ray C. and Jarillo-Herrero, Pablo},
	month = apr,
	year = {2018},
	note = {Publisher: Nature Publishing Group},
	pages = {80--84},
}

@article{lake_pairing_2022,
  title={Pairing symmetry of twisted bilayer graphene: A phenomenological synthesis},
  author={Lake, Ethan and Patri, Adarsh S and Senthil, T},
  journal={Physical Review B},
 doi = {10.1103/PhysRevB.106.104506},
  volume={106},
  number={10},
  pages={104506},
  year={2022},
  publisher={APS}
}

@article{Scheurer2020,
  title = {Pairing in graphene-based moir\'e superlattices},
  author = {Scheurer, Mathias S. and Samajdar, Rhine},
  journal = {Phys. Rev. Res.},
  volume = {2},
  issue = {3},
  pages = {033062},
  numpages = {26},
  year = {2020},
  month = {Jul},
  publisher = {American Physical Society},
  doi = {10.1103/PhysRevResearch.2.033062},
  url = {https://link.aps.org/doi/10.1103/PhysRevResearch.2.033062}
}

@article{Khalaf2022,
  title = {Symmetry constraints on superconductivity in twisted bilayer graphene: Fractional vortices, $4e$ condensates, or nonunitary pairing},
  author = {Khalaf, Eslam and Ledwith, Patrick and Vishwanath, Ashvin},
  journal = {Phys. Rev. B},
  volume = {105},
  issue = {22},
  pages = {224508},
  numpages = {12},
  year = {2022},
  month = {Jun},
  publisher = {American Physical Society},
  doi = {10.1103/PhysRevB.105.224508},
  url = {https://link.aps.org/doi/10.1103/PhysRevB.105.224508}
}

@article{Dynes1984,
  title = {Tunneling Study of Superconductivity near the Metal-Insulator Transition},
  author = {Dynes, R. C. and Garno, J. P. and Hertel, G. B. and Orlando, T. P.},
  journal = {Phys. Rev. Lett.},
  volume = {53},
  issue = {25},
  pages = {2437--2440},
  numpages = {0},
  year = {1984},
  month = {Dec},
  publisher = {American Physical Society},
  doi = {10.1103/PhysRevLett.53.2437},
  url = {https://link.aps.org/doi/10.1103/PhysRevLett.53.2437}
}

@article{Fernandes2013,
  title = {Nematicity as a Probe of Superconducting Pairing in Iron-Based Superconductors},
  author = {Fernandes, Rafael M. and Millis, Andrew J.},
  journal = {Phys. Rev. Lett.},
  volume = {111},
  issue = {12},
  pages = {127001},
  numpages = {5},
  year = {2013},
  month = {Sep},
  publisher = {American Physical Society},
  doi = {10.1103/PhysRevLett.111.127001},
  url = {https://link.aps.org/doi/10.1103/PhysRevLett.111.127001}
}

@article{Fu2014,
  title = {Odd-parity topological superconductor with nematic order: Application to {{Cu$_{x}$Bi$_{2}$Se$_{3}$}}},
  author = {Fu, Liang},
  journal = {Phys. Rev. B},
  volume = {90},
  issue = {10},
  pages = {100509},
  numpages = {5},
  year = {2014},
  month = {Sep},
  publisher = {American Physical Society},
  doi = {10.1103/PhysRevB.90.100509},
  url = {https://link.aps.org/doi/10.1103/PhysRevB.90.100509}
}

@article{Kozii2019,
  title = {Nematic superconductivity stabilized by density wave fluctuations: Possible application to twisted bilayer graphene},
  author = {Kozii, Vladyslav and Isobe, Hiroki and Venderbos, J\"orn W. F. and Fu, Liang},
  journal = {Phys. Rev. B},
  volume = {99},
  issue = {14},
  pages = {144507},
  numpages = {15},
  year = {2019},
  month = {Apr},
  publisher = {American Physical Society},
  doi = {10.1103/PhysRevB.99.144507},
  url = {https://link.aps.org/doi/10.1103/PhysRevB.99.144507}
}

@article{Chichinadze2020,
  title = {Nematic superconductivity in twisted bilayer graphene},
  author = {Chichinadze, Dmitry V. and Classen, Laura and Chubukov, Andrey V.},
  journal = {Phys. Rev. B},
  volume = {101},
  issue = {22},
  pages = {224513},
  numpages = {18},
  year = {2020},
  month = {Jun},
  publisher = {American Physical Society},
  doi = {10.1103/PhysRevB.101.224513},
  url = {https://link.aps.org/doi/10.1103/PhysRevB.101.224513}
}

@article{Hecker2018,
  title={Vestigial nematic order and superconductivity in the doped topological insulator {{Cu$_{x}$Bi$_{2}$Se$_{3}$}}},
  author={Hecker, Matthias and Schmalian, J{\"o}rg},
  journal={npj Quantum Materials},
  volume={3},
  number={1},
  pages={26},
  year={2018},
  publisher={Nature Publishing Group UK London}
}

@article{Jian2021,
  title = {Charge-$4e$ Superconductivity from Nematic Superconductors in Two and Three Dimensions},
  author = {Jian, Shao-Kai and Huang, Yingyi and Yao, Hong},
  journal = {Phys. Rev. Lett.},
  volume = {127},
  issue = {22},
  pages = {227001},
  numpages = {6},
  year = {2021},
  month = {Nov},
  publisher = {American Physical Society},
  doi = {10.1103/PhysRevLett.127.227001},
  url = {https://link.aps.org/doi/10.1103/PhysRevLett.127.227001}
}

@article{Wang2023,
  title={Unusual magnetotransport in twisted bilayer graphene from strain-induced open Fermi surfaces},
  author={Wang, Xiaoyu and Finney, Joe and Sharpe, Aaron L and Rodenbach, Linsey K and Hsueh, Connie L and Watanabe, Kenji and Taniguchi, Takashi and Kastner, M. A. and Vafek, Oskar and Goldhaber-Gordon, David},
  journal={Proceedings of the National Academy of Sciences},
  volume={120},
  number={34},
  pages={e2307151120},
  year={2023},
  publisher={National Academy of Sciences}
}

@article{Fernandes_Fu,
  title = {Charge-$4e$ Superconductivity from Multicomponent Nematic Pairing: Application to Twisted Bilayer Graphene},
  author = {Fernandes, Rafael M. and Fu, Liang},
  journal = {Phys. Rev. Lett.},
  volume = {127},
  issue = {4},
  pages = {047001},
  numpages = {6},
  year = {2021},
  month = {Jul},
  publisher = {American Physical Society},
  doi = {10.1103/PhysRevLett.127.047001},
  url = {https://link.aps.org/doi/10.1103/PhysRevLett.127.047001}
}

@article{Sigrist1991,
  title = {Phenomenological theory of unconventional superconductivity},
  author = {Sigrist, Manfred and Ueda, Kazuo},
  journal = {Rev. Mod. Phys.},
  volume = {63},
  issue = {2},
  pages = {239--311},
  numpages = {0},
  year = {1991},
  month = {Apr},
  publisher = {American Physical Society},
  doi = {10.1103/RevModPhys.63.239},
  url = {https://link.aps.org/doi/10.1103/RevModPhys.63.239}
}

@article{Vaglio1993,
  title = {Explanation of the resistance-peak anomaly in nonhomogeneous superconductors},
  author = {Vaglio, R. and Attanasio, C. and Maritato, L. and Ruosi, A.},
  journal = {Phys. Rev. B},
  volume = {47},
  issue = {22},
  pages = {15302--15303},
  numpages = {0},
  year = {1993},
  month = {Jun},
  publisher = {American Physical Society},
  doi = {10.1103/PhysRevB.47.15302},
  url = {https://link.aps.org/doi/10.1103/PhysRevB.47.15302}
}

@article{Sprau2017,
  title={Discovery of orbital-selective Cooper pairing in {Fe}{Se}},
  author={Sprau, Peter O and Kostin, Andrey and Kreisel, Andreas and B{\"o}hmer, Anna E and Taufour, Valentin and Canfield, Paul C and Mukherjee, Shantanu and Hirschfeld, Peter J and Andersen, Brian M{\o}ller and Davis, JC S{\'e}amus},
  journal={Science},
  volume={357},
  number={6346},
  pages={75--80},
  year={2017},
  publisher={American Association for the Advancement of Science}
}

@article{Nag2025,
  title={Highly anisotropic superconducting gap near the nematic quantum critical point of {Fe}{Se}$_{1-x}${S}$_x$},
  author={Nag, Pranab Kumar and Scott, Kirsty and de Carvalho, Vanuildo S and Byland, Journey K and Yang, Xinze and Walker, Morgan and Greenberg, Aaron G and Klavins, Peter and Miranda, Eduardo and Gozar, Adrian and others},
  journal={Nature Physics},
  volume={21},
  number={1},
  pages={89--96},
  year={2025},
  publisher={Nature Publishing Group UK London}
}

@article{Dodaro2018,
  title = {Phases of a phenomenological model of twisted bilayer graphene},
  author = {Dodaro, J. F. and Kivelson, S. A. and Schattner, Y. and Sun, X. Q. and Wang, C.},
  journal = {Phys. Rev. B},
  volume = {98},
  issue = {7},
  pages = {075154},
  numpages = {8},
  year = {2018},
  month = {Aug},
  publisher = {American Physical Society},
  doi = {10.1103/PhysRevB.98.075154},
  url = {https://link.aps.org/doi/10.1103/PhysRevB.98.075154}
}

@article{Wu2017,
  title={Spontaneous breaking of rotational symmetry in copper oxide superconductors},
  author={Wu, Jie and Bollinger, AT and He, X and Bo{\v{z}}ovi{\'c}, I},
  journal={Nature},
  volume={547},
  number={7664},
  pages={432--435},
  year={2017},
  publisher={Nature Publishing Group UK London}
}

@article{Howald2003,
  title={Coexistence of periodic modulation of quasiparticle states and superconductivity in {Bi}$_2${Sr}$_2${Ca}{Cu}$_2${O}$_{8+\delta}$},
  author={Howald, C and Eisaki, H and Kaneko, N and Kapitulnik, A},
  journal={Proceedings of the National Academy of Sciences},
  volume={100},
  number={17},
  pages={9705--9709},
  year={2003},
  publisher={National Academy of Sciences}
}

@article{Fernandes2022,
  title={Iron pnictides and chalcogenides: a new paradigm for superconductivity},
  author={Fernandes, Rafael M and Coldea, Amalia I and Ding, Hong and Fisher, Ian R and Hirschfeld, PJ and Kotliar, Gabriel},
  journal={Nature},
  volume={601},
  number={7891},
  pages={35--44},
  year={2022},
  publisher={Nature Publishing Group UK London}
}

\section*{Author Contributions Statement}
S.S. and J.M.P. fabricated devices, conducted measurements, and performed data analysis. G.P. and R.M.F. performed theoretical calculations. K.W. and T.T. provided hBN samples. J.M.P. and P.J.-H. supervised the project. S.S., G.P., R.M.F., P.J.-H., and J.M.P. discussed the results and wrote the manuscript with contributions from all authors.

\section*{Correspondence}
Correspondence should be sent to J.M.P. (parkjane@mit.edu) and P.J-H. (pjarillo@mit.edu).

\section*{Competing Interests Statement}
The authors declare no competing interests.

\end{document}